\documentclass[%
 reprint,
 amsmath,amssymb,
 aps,
]{revtex4-1}

\usepackage{graphicx}
\usepackage{dcolumn}
\usepackage{bm}
\usepackage{xcolor}
\usepackage{soul}

\begin{document}

\preprint{APS/123-QED}

\title{
Anisotropic nano-scale resolution in 3D Bragg coherent diffraction imaging
}

\author{Mathew J. Cherukara}
\author{Wonsuk Cha}
\author{Ross J. Harder}%
 \email{rharder@aps.anl.gov}
\affiliation{%
 Advanced Photon Source, Argonne National Laboratory, Lemont, IL 60439, USA\\
}%

\date{\today}

\begin{abstract}
We demonstrate that the resolution of three-dimensional (3D) real-space images obtained from Bragg x-ray coherent diffraction measurements is direction dependent. We propose and demonstrate the effectiveness of a metric to determine the spatial resolution of images that accounts for the directional dependence. The measured direction dependent resolution of $\sim4-9nm$ nm is about 2 times higher than the best previously obtained 3D measurements. Finally, we quantify the relationship between the resolution of recovered real-space images and dosage, and discuss its implications in the light of next generation synchrotrons.


\end{abstract}

\pacs{Valid PACS appear here}
\maketitle


Bragg coherent x-ray diffraction imaging (BCDI) is a powerful x-ray imaging technique that  provides 3D structural and strain information at a few nanometer spatial resolution under \textit{operando} conditions\cite{Robinson2009}. Consequently, the technique has been used to provide valuable insight into various mechanistic and chemical processes such as dissolution and crystallization\cite{Clark2015b}, transient melting\cite{Clark2015}, and phonon propagation\cite{Cherukara2017,Ulvestad2017BraggTimescales}. In this technique, intensities of scattered coherent x-rays are measured in the far-field about a Bragg peak\cite{Pfeifer2006}. Iterative algorithms that apply certain mathematical constraints in real and reciprocal space are used to recover the real space object and lost reciprocal space phase\cite{Harder2010}. As BCDI is a lensless technique, the achievable image resolution is not optic limited, and is in principle only limited by the wavelength of x-ray which is typically of the order of 1 {\AA}\cite{Miao2015}. Though Angstrom resolution imaging has not yet been approached with x-rays, there is research to support that this may be achieved with next generation synchrotrons\cite{Dietze2015CoherentResolution}.  To date, the best reported resolution from a BCDI measurement in 3D is 8 nm along a preferential direction\cite{Clark2012}.  Modern scanning coherent diffraction techniques have also been applied to Bragg geometry and have revealed features with 6 nm structure\cite{Labat2015InversionImaging}.  In the small angle scattering case, there has been 3 nm features observed in 2D and better than 10 nm resolution in 3D via tomography combined with CDI\cite{Takahashi2010df}.  Also, recent efforts to extrapolate the resolution beyond that enabled by the finite detector size have shown promise\cite{Laty2013,Latychevskaia2015ImagingPatterns}.

Central to the present discussion is the concept of image resolution in coherent imaging.  Traditionally, a microscopes image resolution can be quantified using the optical transfer function (OTF) of the lens system.  Defined as the Fourier Transform of the point spread function of the imaging system, it is loosely analogous to the mechanism by which high resolution is generated in CDI.  A broad OTF means that high spatial frequencies are transmitted through the imaging system, which then correspond to short distances in the image.  The resolution of a coherent diffraction image comes from the fact that signal is measured to high spatial frequency in reciprocal space, identically to a microscope with a broad OTF.  The image resolution is therefore determined by the usable signal at large spatial frequency, or momentum transfer of the photons. In coherent imaging methods, the image is produced by computationally recovering the phases of the scattered waves and inverting to a direct space image via Fourier Transform.  The phased image in BCDI is also generated computationally, but the data is measured about a Bragg peak of the lattice by rotating the sample through a small angular rocking curve to acquire the 3D reciprocal space volume.  Resolution is improved by measuring longer to allow signal to accumulate at large momentum transfer relative to the reciprocal space point defining the Bragg peak. 

In this work, we report spatial resolution of $\sim$4-9 nm, that is dependent on the direction in 3D. Furthermore, we propose and show the effectiveness of a new metric for resolution that fully elucidates the directional nature of coherent x-ray scattering. We note that while we have performed the measurements and subsequent analysis in the Bragg geometry, the methodology described in this work is trivially transferred to any coherent diffraction method or sample state\cite{Kirian2015DirectLaser}, such as single particle imaging\cite{vanderSchot2015ImagingLaser}.


The BCDI measurement was performed at beamline 34-ID-C of the Advanced Photon Source (APS). Isolated gold (Au) nanocrystals on a silicon (Si) substrate, obtained by dewetting a Au thin film on Si, was placed at the center of the diffractometer. A Si (111) monochromator was used to set the photon energy of the x-ray beam to 9.0 keV. The diffraction signal was collected using a Timepix detector (Amsterdam Scientific Instruments) with 512x512 pixels each of which is 55x55 $\mu$m$^2$. To obtain the 3D coherent diffraction pattern about a (111) Bragg peak, the detector was placed at the appropriate $2\theta$, and the sample stage was rocked  through an angle of 2 degrees in steps of 0.008 degrees with an exposure time of 2 s per point on the rocking curve. The same scan was repeated 25 times, after centering the crystal in the x-ray beam between scans to correct for any instrumentation drift. The diffraction data used in the following analysis was obtained by summing the 25 scans, after correcting for minor shifts in the position the Bragg peak on the detector.

	A common technique to measure the spatial resolution in CDI is the radially averaged phase retrieval transfer function (PRTF)\cite{Chapman2006a,Ravasio2009,Takahashi2009}. Analogous to the OTF, the width of the radially averaged PRTF, is a measure of reconstructed object resolution. To obtain the PRTF, multiple images are obtained from the data, with different initial, random guesses for the phase. The PRTF then, is the ratio of the average diffraction amplitude obtained from the images to the measured amplitudes, and is a measure of how reliably the phases are recovered. The radially averaged PRTF is obtained by averaging this ratio over shells of constant scattering vector magnitude (q):
\begin{equation}
PRTF(q)=\frac{|<\Gamma_{i}(q)>|}{\sqrt{I_{m}(q)}}
\end{equation}
where $\Gamma_i$ is the diffraction amplitude corresponding to the $i^{th}$ phased image and $I_m$ is the measured intensity. Hence, where the phases are consistently recovered, typically at low q, the PRTF is $\sim$1. With increasing q, the consistency of the recovered phases drops (PRTF$<1$), and the image resolution is defined as the point at which PRTF$<1/e$ where $e$ is Euler's number. Figure 1 shows the PRTF obtained by averaging images obtained from 10 different starting guesses. Figure 1(a) shows a 3D rendering of the PRTF with the contour level set to $0.37\sim 1/e$\cite{Chapman2006a}, while Fig. 1 (b)-(d) show slices through the PRTF where contour lines in white represent $PRTF\sim 1/e$\cite{Chapman2006a}. Figure 1(e) shows the real space image of the crystal. 
As evinced by the complex shape of the PRTF, the resolution is extremely sensitive to direction, and a radially averaged number for the spatial resolution is mis-representative of the true nature of the image resolution. For instance, the resolution along the direction normal to largest facets of the crystal (along Y) is $\sim 230 \: \mu m^{-1}$ ($\sim 4.3 \: nm$) (Fig. 1(d) and (e)), while in in-plane directions, the resolution varies from $\sim 40 \: \mu m^{-1}$ ($\sim 25 \: nm$) to $\sim 80 \: \mu m^{-1}$ ($\sim 12.5 \: nm$) (Fig. 1(c)). Furthermore, by definition, the PRTF is a measure of how well the reconstructed images compare with each other, not how well they compare with a global reference. Hence, if the phased images always stagnate in the same local minima during the iterative phase retrieval, the PRTF will suggest an exaggerated resolution\cite{Dietze2015CoherentResolution}.
\begin{figure}
\includegraphics[width=0.45\textwidth]{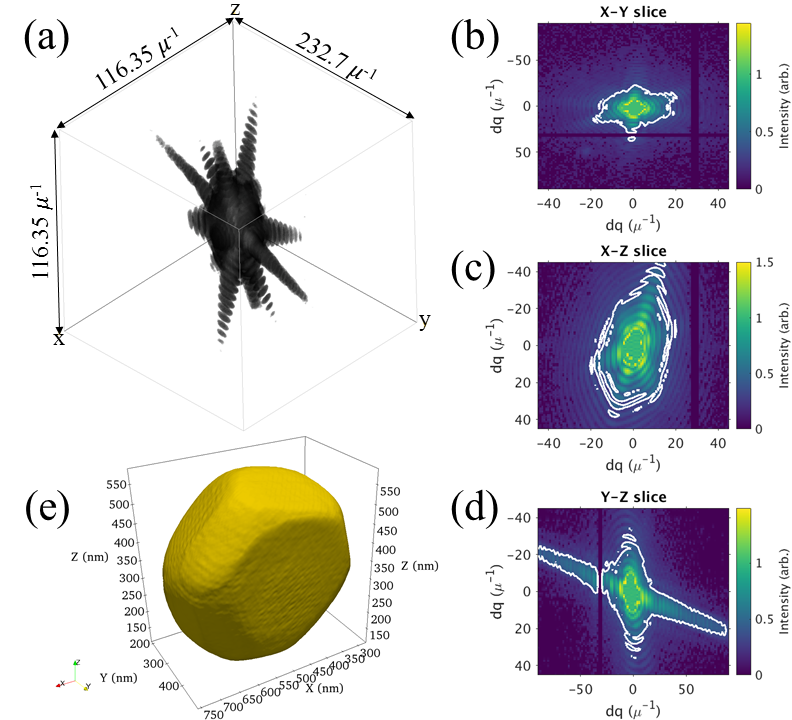}
\caption{(a) 3D rendering of the phase retrieval transfer function with the contour level set at $0.37\sim1/e$. (b-d) Slices through the transfer function showing the 0.37 contour lines. (e) shows the crystal image with the surface contour set to 0.3 of the maximum density.}
\end{figure}

To address the shortcomings of the PRTF described in the previous paragraph, we propose a metric that defines the resolution in real space as opposed to reciprocal space. This provides the convenience of working in the laboratory coordinate system after the appropriate coordinate transformation has been employed to go from the detector frame, to the laboratory frame\cite{Leake2010a}. We define the resolution as the width of the spreading function that blurs an ideal, perfectly sharp crystal boundary to the actual, recovered crystal shape and density. To obtain this ideal crystal structure, we first define the surface of the crystal at a specific fraction of the maximum real space intensity (say 0.3 of the max).  This essentially defines the contrast at which we are computing the resolution of the image.  The image is then binarized to this threshold, i.e, voxels that have intensity$\geq$0.3 are set to 1, while the remaining voxels are set to 0. Our goal then is to obtain the blurring function which when convolved with this idealized crystal, gives the actual image. To extract this blurring function, we employ the Richardson-Lucy (RL) deconvolution algorithm\cite{Richardson1972Bayesian-BasedRestoration,Lucy1974AnDistributions}. RL deconvolution has been widely used to improve image quality in magnetic resonance imaging (MRI) scans\cite{DellAcqua2010ADeconvolution}, reducing motion blur\cite{Tai2011Richardson-LucyPath} and even coherent diffraction imaging with a partially coherent beam\cite{Clark2012}. To obtain the blurring function $b$ from the perfect crystal $P$ and actual crystal $A$, so that $P\otimes b=A$, RL deconvolution solves the following equation iteratively:
\begin{equation}
b^{(t+1)}=b^t.(\frac{A}{b^t\otimes P}\otimes \hat{P})
\end{equation}
where $\hat{P}$ is the flipped perfect crystal, such that $\hat{P}_{nml}=P_{(i-n)(j-m)(k-l)},0\le n,m,l\le i,j,k$. It has been shown that if this iterative procedure converges, it converges to the maximum likelihood solution for $b$\cite{Shepp1982MaximumTomography}.  We define the deconvolution error $\epsilon$ as the normalized difference between the current and previous iterate, 
\begin{equation}
\epsilon=\frac{|b^{t+1}-b^t|}{\sum\limits_{i,j,k} b^t_{ijk}}
\end{equation}

\begin{figure}
\includegraphics[width=0.47\textwidth]{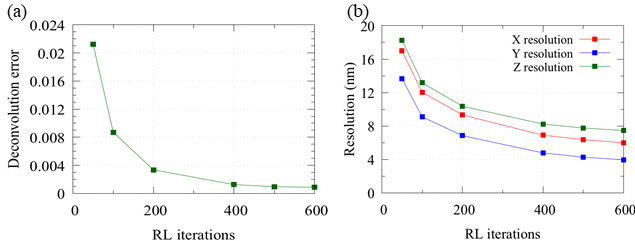}
\caption{(a) Error in the deconvolved blurring function as a function of number of iterations of the Richardson-Lucy (RL) algorithm. (b) Calculated resolution in the lab X, Y and Z directions as a function of the number of RL iterations. The convergence in calculated resolution closely mirrors the convergence in the deconvolution error.}
\end{figure}

Figure 2(a) shows the convergence in the deconvolution error $\epsilon$ as a function of number of iterations. Finally, we define the resolution of the image along a given direction as the full width half max (FWHM) of the blurring function in that direction, which we estimate by fitting a Gaussian. Figure 2(b) shows the calculated resolution in laboratory X, Y, and Z directions as a function of iteration number. Evidently, the convergence in the calculated resolution closely mirrors the convergence in the deconvolution error.

\begin{figure*}
\includegraphics[width=0.95\textwidth]{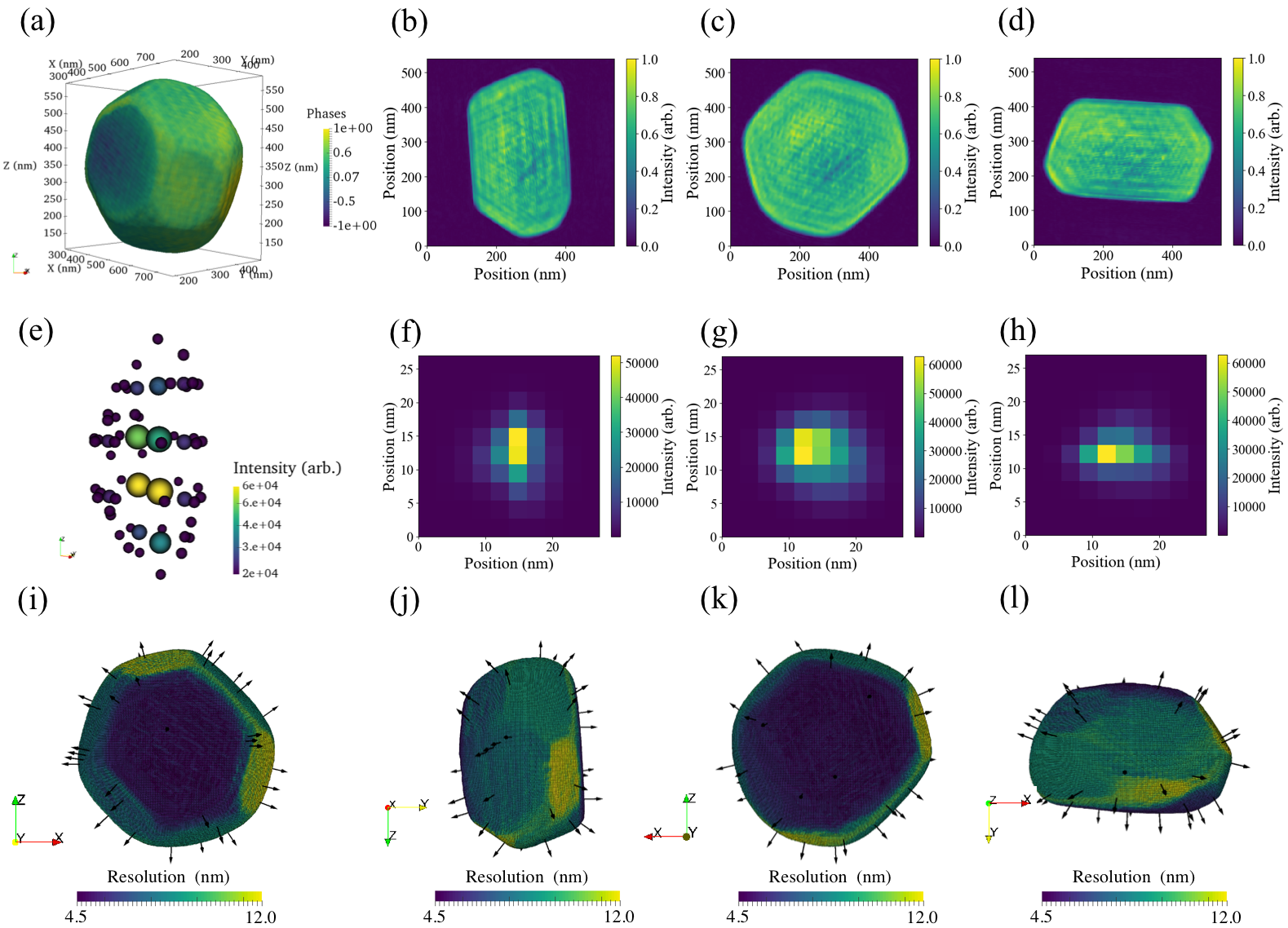}
\caption{(a) Image of an Au nanocrystal showing sharp $\{111\}$ and $\{001\}$ facets. Coloring is by the recovered phase. (b-d) Slices through the Y-Z, X-Z and X-Y planes showing the amplitude variation through the crystal. (e) Computed 3D blurring function. Color and size of sphere correspond to the intensity. (f-h) Slices through the blurring function along the Y-Z, X-Z and X-Y planes. The extremely narrow spreading function that is recovered ($\sim$1-2 pixels wide) is representative of the high resolution obtained. (i-l) Points on the 0.3 isosurface coloured by the resolution along the direction normal to the crystal surface at that point. Black arrows denote the direction of the surface normal at select points. The top and bottom facets of the crystal (Fig.3 (i,k)) show the lowest resolution.}
\end{figure*}

We next turn our attention to the shape of the computed 3D blurring function. Figure 3(a) shows the image of the crystal colored by the recovered phase, with a contour level of 0.3 used to define the crystal surface. Clearly defined large facets corresponding to the {111} and {001} planes along with narrow {110} facets can be identified, as predicted by the Wulff construction\cite{Barmparis2012DependenceTheory}. Figure 3(b)-(d) show 2D slices through the crystal, coloring is by amplitude. Figure 3(e) shows a 3D render of the corresponding blurring function, where point color and size are representative of the magnitude. The convolution of this recovered 3D blurring function in Fig. 3(e) with the actual crystal shape gives the imaged crystal in Fig. 3(a). Figure 3(f)-(g) show 2D slices through the blurring function, where coloring is again by magnitude. We first note the extremely sharp blurring function that is obtained. Indeed, the width of the blurring function is less than 2 pixels in the laboratory Y direction, where the resolution is $\sim$4nm. We also observe that the shape of the blurring function correlates with the shape of the crystal; the resolution in directions normal to crystal facets is better than other directions.

We can extend this analysis to any arbitrary direction. In particular, it is useful to calculate the resolution in directions normal to the crystal facets. Figure 3(i)-(l) show the crystal structure with every point on the surface colored by the resolution in the direction along the surface normal at that point. Black arrows indicate the direction of the surface normal at that point. For instance, Fig.s 3(i,k) show a view of the top and bottom surfaces of the crystal ({111} facets), where the resolution calculated along the normals to the surface is $\sim 4.5nm$, which explains the extremely high resolution calculated in the lab Y direction (Fig. 2). This is also consistent with the calculated resolution using the PRTF, which yielded a value of $\sim$4.3 nm. This extremely high resolution is a consequence of this particular facet pair ({111} facets along +Y and -Y) having a significantly larger surface area, which contributes strongly to scattering in that direction. Interestingly, for two facets of comparable area, the directional sensitivity appears to have a crystallographic bias, with {100} facets showing lower resolution (Fig.s 3(i-k)), possibly due to lower packing density on {100} surfaces compared to {111} surfaces.

The procedure described previously provides a means of quantifying the resolution obtained from BCDI measurements, and in particular of quantifying the \textit{directionality} of the resolution. In our final discussion, we employ this method to calculate the dependence of spatial resolution on the exposure time. Figure 4 shows the calculated resolution in laboratory X, Y, and Z directions as a function of exposure time. To obtain diffraction data with different cumulative exposure times, a subset of the acquired scans are summed together. As seen in Fig. 4 (a), the calculated resolution in lab Y, which corresponds to the direction normal to the largest face of the crystal (Fig. 3), is considerably worse when fewer scans are included in the phased data. The scattering intensity with increasing scattering vector ($\vec{q}$) is predicted to decay as $q^{(-4)}$\cite{Schroer2008}, and consequently, the required dose (D) for a given theoretical resolution (r) is given by $D=a*r^{-4}$, where a is a constant of proportionality\cite{Marchesini2003CoherentLimitations}. Assuming the dose scales linearly with exposure time, the maximum achievable resolution for a given exposure time (t) is given by,
\begin{equation}
d=c*t^{-0.25}
\end{equation}
where c is a proportionality constant. However, as described above, the resolution of CDI is not necessarily determined by the maximum $\vec{q}$ at which there is scattered intensity, but by the maximum $\vec{q}$ with accurately phased intensity. Therefore it is not obvious that the image resolution in real space should follow the same power law as the scattered intensity in reciprocal space. Fig. 4 (b) shows power law fits to the resolution as a function of exposure time. Fitted exponents shown in the plot are $\sim$0.25, suggesting that indeed the resolution of the  images does scale in the same manner as the scattered intensity in reciprocal space. This power law for the dependence of image resolution on dose is also in agreement with simulations of Starodub et al.\cite{Starodub2007}.

\begin{figure}
\includegraphics[width=0.47\textwidth]{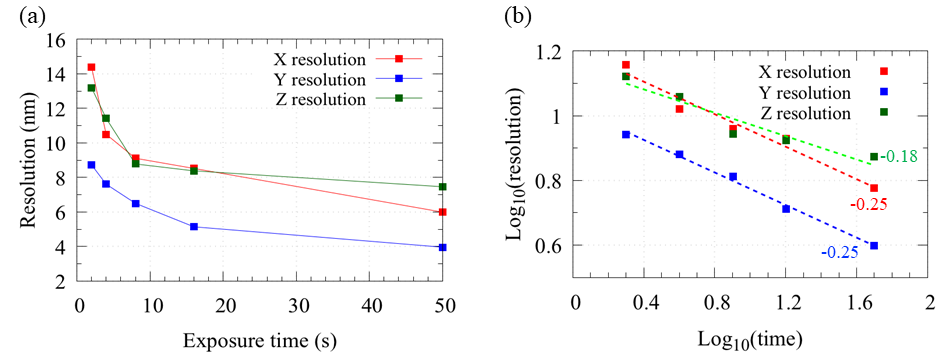}
\caption{(a,b) Calculated resolution in lab X, Y, and Z as a function of exposure time per point on the rocking curve. Dashed lines in (b) show linear fits on the log scale, with the fit slopes shown in colored text.}
\end{figure}

To conclude, we have shown that the resolution of the images obtained from BCDI measurements is direction dependent. To account for this directionality, we have described a new metric to calculate the resolution, that agree well with calculations from the PRTF. Using our new metric, we evaluated the directionally dependent spatial resolution of BCDI of a gold nanocrystal.  The measured resolution for the same crystal varies from $~4 nm$ to $~ 9nm$, with the highest resolution being in the direction normal to the surface of the large (111) facet that is perpendicular to the substrate. Finally, we evaluate the x-ray exposure time dependence of the resolution function of images.  We find that the resolution of retrieved images scales with the exposure time as $t^{-4}$, which mirrors the power law describing the scattering intensity at given $\vec{q}$. This suggests that the iterative retrieval algorithms, if converged, do not affect the scaling between the measured intensity and maximum achievable resolution.

The proposed metric for resolution has the advantage of being based on the real space images, and as such, the calculated directional values are in the laboratory frame. Additionally, PRTF, where tens or hundreds of phased images are necessary to obtain the resolution, by using the method proposed herein, resolution can be directly determined from a single phase retrieval run. This 10x-100x reduction in compute time will be particularly advantageous in the light of facility upgrades such as the proposed APS Upgrade project, where the acquired data sets, and time required to run iterative phasing, are expected to increase dramatically.

A natural ramification of the analysis presented in this article is that image resolution is dependent on the sample; it's size and morphology. Consequently, we believe that rapid determination of the anisotropic image resolution during in-situ BCDI experiments will enable real-time understanding of structural changes occurring in a sample.  Frequently experiments are conducted and images are evaluated with little or no feedback as to whether the changes seen are indeed visible at the resolution of the obtained image or are simply features at the noise level of the measurement\cite{Ulvestad2016InNanoparticles}.  With real-space deconvolution of the image resolution, one can determine immediately if more data is required at a given state of the sample.

This work was supported by Argonne LDRD 2018-019-N0:  A.I C.D.I: Atomistically Informed Coherent Diffraction Imaging. Use of the Advanced Photon Source was supported by the U. S. Department of Energy, Office of Science, Office of Basic Energy Sciences, under Contract No. DE-AC02-06CH11357.

\bibliography{Mendeley_BCDIResolution_v2.bib}

\end{document}